\batchmode
\makeatletter
\def\input@path{{/home/yecheng/IV2019/}}
\makeatother
\documentclass[10pt,twocolumn,english]{ieeeconf}
\usepackage[T1]{fontenc}
\usepackage[latin9]{inputenc}
\usepackage{geometry}
\usepackage{color}
\usepackage{float}
\usepackage{textcomp}
\usepackage{amsmath}
\usepackage{amssymb}
\usepackage{graphicx}
\usepackage{setspace}

\makeatletter

\newcommand{\lyxmathsym}[1]{\ifmmode\begingroup\def\b@ld{bold}
  \text{\ifx\math@version\b@ld\bfseries\fi#1}\endgroup\else#1\fi}

\providecommand{\tabularnewline}{\\}
\floatstyle{ruled}
\newfloat{algorithm}{tbp}{loa}
\providecommand{\algorithmname}{Algorithm}
\floatname{algorithm}{\protect\algorithmname}

\usepackage{cite}

\usepackage{babel}
\usepackage{subfig}
\usepackage{caption}

\makeatother

\usepackage{babel}
\begin{document}
\title{\pagenumbering{gobble}\vspace{0.0in}An Interactive LiDAR to Camera
Calibration}
\author{Yecheng Lyu, Lin Bai, Mahdi Elhousni and Xinming Huang\thanks{This work is partially supported by U.S. NSF Grant CNS-1626236.}\textcolor{black}{{}
}\thanks{All authors are with Department of Electrical and Computer Engineering,
Worcester Polytechnic Institute, Worcester, MA 01609, USA. Email:
\{ylyu,lbai2, melhousni, xhuang\}@wpi.edu}}
\maketitle
\begin{abstract}
\textcolor{black}{Recent progress in the automated driving system
(ADS) and advanced driver assistant system (ADAS) has shown that the
combined use of 3D light detection and ranging (LiDAR) and the camera
is essential for an intelligent vehicle to perceive and understand
its surroundings. LiDAR-camera fusion requires precise intrinsic and
extrinsic calibrations between the sensors. However, due to the limitation
of the calibration equipment and susceptibility to noise, algorithms
in existing methods tend to fail in finding LiDAR-camera correspondences
in long-range. In this paper, we introduced an interactive LiDAR to
camera calibration toolbox to estimate the intrinsic and extrinsic
transforms. This toolbox automatically detects the corner of a planer
board from a sequence of LiDAR frames and provides a convenient user
interface for annotating the corresponding pixels on camera frames.
Since the toolbox only detects the top corner of the board, there
is no need to prepare a precise polygon planar board or a checkerboard
with different reflectivity areas as in the existing methods. Furthermore,
the toolbox uses genetic algorithms to estimate the transforms and
supports multiple camera models such as the pinhole camera model and
the fisheye camera model. Experiments using Velodyne VLP-16 LiDAR
and Point Grey Chameleon 3 camera show robust results.}
\end{abstract}

\section{\textcolor{black}{Introduction}}

\textcolor{black}{Automated driving systems (ADS) and advanced driver
assistant systems (ADAS) equipped on intelligent vehicles rely on
multiple sensors to perceive their surroundings. In recent research
works, LiDAR-based algorithms have shown their advantage on drivable
region segmentation\cite{lyu2018real_time} \cite{lyu2018chipnet},
object detection\cite{yan2018second}, and simultaneous localization
and mapping \cite{zhang2014loam} \cite{shan2018lego}. LIDARs are
also fused with cameras to improve the accuracy of 3D object detection
\cite{chen2017multi}. However, calibration between LiDAR and camera
devices is required to assign the detections to the same coordinate
frames so that we can fuse the sensor data. Owing to the development
of Multiple View Geometry and computer vision, the models of 3D to
2D projection has been well established. Nevertheless, limitations
are observed in existing calibration algorithms when applied to LiDAR-camera
systems on intelligent vehicles.}

Figure 1 shows our autonomous vehicle prototype. The LiDAR-camera
system is designed to detect up to 100 meters in front of the vehicle.
But most of the existing LiDAR-to-camera calibration algorithms are
proposed for indoor use and are validated in meters rage\cite{Geiger2012}\cite{Pereira2016}\cite{Park2014}.
As shown in Figure \ref{fig:Resolution-of-LiDAR}, the increase of
sensing range leads to lower resolution of the camera lens system
and larger offset on LiDAR points when projected to the image plane.
Thus calibration of the LiDAR-to-camera system needs to be improved.

\begin{figure}
\begin{centering}
\includegraphics[width=8cm]{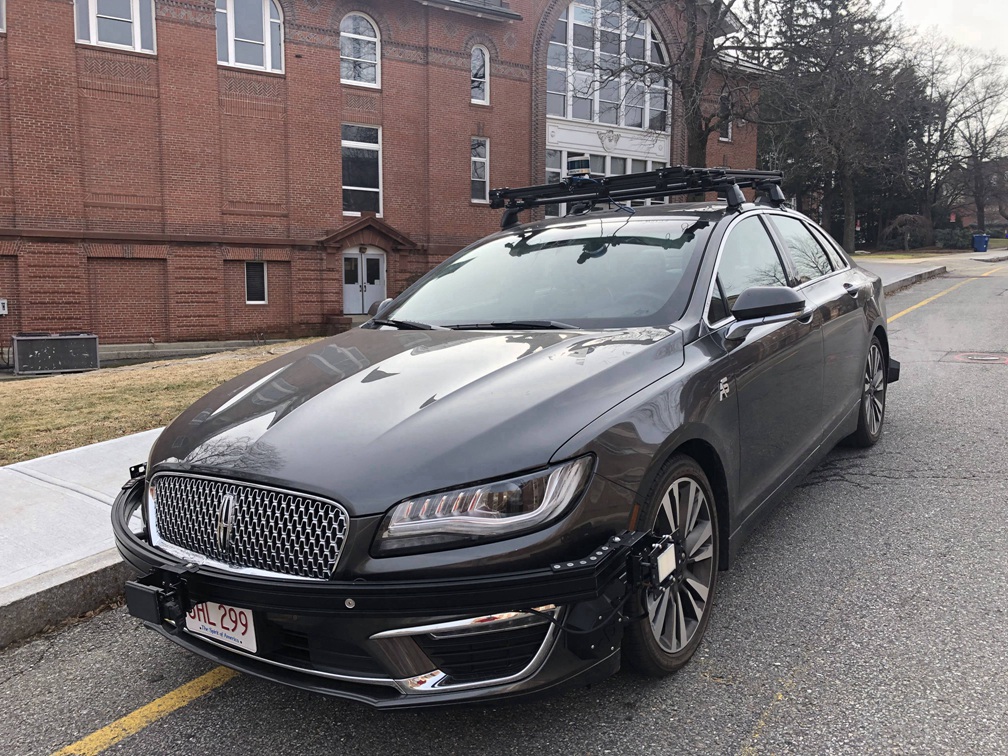}
\par\end{centering}
\caption{A Lincoln MKZ equipped with 1 LiDAR, 1 front camera and 10 side cameras}
\end{figure}

In addition, LiDAR sensors have limited resolution according to their
vertical axis, which makes it difficult to find correspondences between
the points in a LiDAR frame and the exact pixels in the corresponding
camera frame directly. Previous works have proposed several indirect
methods to estimate the correspondences. Rangesh et al \cite{Rangesh2017}
and Geiger et al \cite{Geiger2012} assumed that the checkerboards
have a flat surface and that all the LiDAR points on each checkerboard
should be co-planar. However, traditional checkerboards made from
cardboard, wood or aluminum are not flat enough for outdoor LiDAR-camera
system calibration. To overcome this issue, \cite{Rangesh2017} used
a glass made calibration board to ensure near-perfect flatness and
rigidity. On the other hand, Park et al \cite{Park2014} and Pereira
et al \cite{Pereira2016} employed objects in special shapes to estimate
the point-to-point correspondences. However, their works were also
sensible to the lack of rigidity. Due to the same reasons, target-less
approaches \cite{Irie2016} also experienced large projection offsets.
Moreover, due to the low vertical resolution, labeling correspondences
for long-range calibration is difficult because the LiDAR sensor may
not get enough lines of scans on the checkerboards or target objects.
Thus few of the existing works can calibrate and validate their results
for a distance longer than 5 meters.

In this paper, we introduce a new toolbox for LiDAR to camera calibration
using a polygonal board. Through the automatic detection of the vertices
in LiDAR frames and manual labeling the correspondence in camera frames,
our solution collects direct point-to-point correspondences between
LiDAR and camera coordinates. The correspondence pairs are used to
estimate the intrinsic and extrinsic transforms via a genetic algorithm
based approach. The rest of this paper is organized as follows: Section
\ref{sec:Related-work} discusses the related works. Section \ref{sec:Methodology}
describes the calibration models in the proposed method. In Section
\ref{sec:Experiment}, we evaluate the proposed method on our LiDAR
and Camera recordings. Section \ref{sec:Conclusions-and-future} gives
the conclusions.

\begin{figure}
\begin{centering}
\includegraphics[width=8cm]{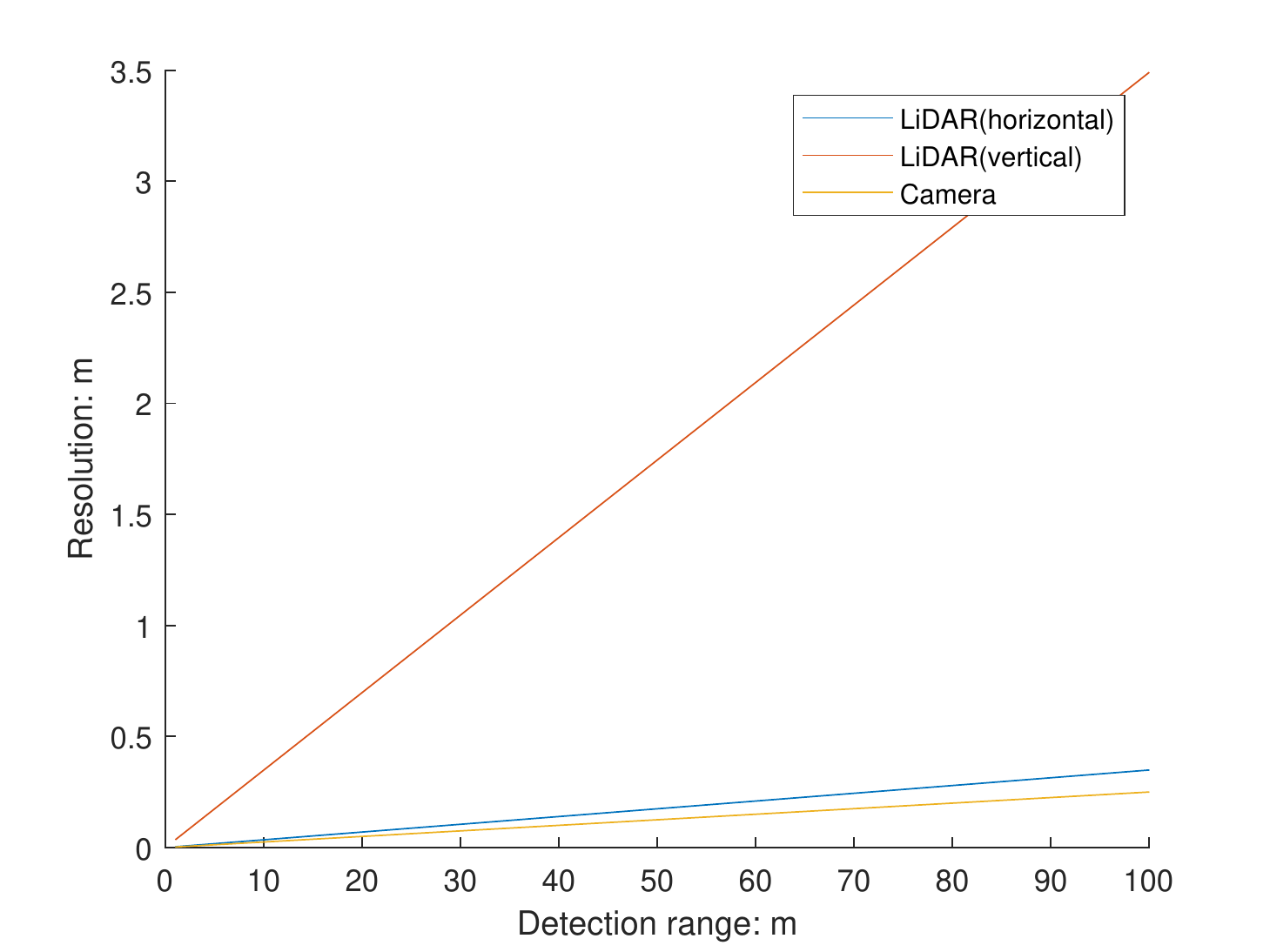}
\par\end{centering}
\caption{Resolution of LiDAR and camera. The LiDAR has resolution of $0.2\lyxmathsym{\protect\textdegree}$
in vertical and $2\lyxmathsym{\protect\textdegree}$ in horizontal.
The scale/focus ratio of the camera is 400.\label{fig:Resolution-of-LiDAR} }

\end{figure}

\section{\textcolor{black}{Related Work\label{sec:Related-work}}}

\subsection{LiDAR-camera correspondence collection}

Acquiring the LiDAR-camera correspondence is the first step of calibration.
Since point-to-point correspondences are difficult to measure, two
categories of indirect approaches were proposed in the literature.

one method is to estimate the point-to-point correspondences via knowledge
of specific objects. \cite{Park2014} estimated the vertices of the
rigid checkerboards in LiDAR frames by detecting the checkerboard
edges. The approach of \cite{Pereira2016} detected the surface of
the target ball and estimated the position of its center. This method
relies on the clarity and precision of the target objects. They also
assume that an intrinsic calibration of the camera is valid.

The other method does not rely on point-to-point correspondences.
\cite{Geiger2012} and \cite{Rangesh2017} assumed the LiDAR points
on the checkerboard are coplanar and performed calibrations through
planar-to-planar correspondences. Ishikawa et al \cite{Ishikawa2018}
estimated the motion of LiDAR and camera separately and determined
the intrinsics and extrinsics  through motion-to-motion correspondences.
Banerjee et al \cite{Banerjee2018} detected the edges of objects
in camera frames and calibrated through edge-to-edge correspondences.
These approach s not sensitive to the quality of calibration equipment
but relies on the segmentation of LiDAR and camera frames.

In this paper, we propose a point-to-point correspondence based approach.
To overcome the limitations, we detect the vertices of target objects
in a sequence of frames rather than a single frame. We also involve
manual annotation of vertices in camera frames to increase the accuracy
of correspondences.

\subsection{Solver of calibration }

Different optimizers have been used to estimate the parameters in
the calibration models. Hulik et al \cite{Hulik2014} used 3D Hough
transform to search for a continuous plane in point clouds. Velas
et al \cite{Velas} and Vasconcels et al \cite{Vasconcelos} applied
the RANSAC method to solve the extrinsic calibration. Paynot et al
\cite{Peynot2010} and Zhang \cite{Zhang2008} introduced a probabilistic
model to estimate the likelihood of a transformation and then iterated
using the Levenberg\textendash Marquardt algorithm. Heikkila et al
\cite{Heikkila1997} also used the Levenberg\textendash Marquardt
algorithm but applied direct linear transform (DLT) to solve the initial
estimation of camera intrinsic parameters. Pandey et al \cite{Pandey2015}
estimated the extrinsic calibration based upon the maximization of
mutual information. In our work, a genetic algorithm is applied to
estimate the parameters in extrinsic and intrinsic transforms since
it works for non-linear models and avoids local optimal.

\section{\textcolor{black}{Methodology\label{sec:Methodology}}}

\textcolor{black}{This section gives an overview of our work on LiDAR
to camera calibration. In this section, we first demonstrate the models
applied in the proposed calibration, describe the data collection
and processing, and then present the use of the genetic algorithm
to solve the calibration task.}

\subsection{Calibration models}

Models of LiDAR to camera projection have been well investigated.
Suppose we have a scanned point $(x,y,z)$ in LiDAR coordinates, its
corresponding point $(u,v,w)$ in camera coordinates, and its corresponding
pint $(i,j)$ in the image plane. The transformation from LiDAR to
the image includes two parts, as described in \cite{Heikkila1997}\cite{Geiger2012}.

The first one is the extrinsic transformation that is the projection
model from LiDAR to camera coordinate. This 6-DOF matrix can be expressed
as (\ref{eq:extrinsic}).

\begin{doublespace}
\begin{equation}
\begin{array}{c}
\left[\begin{array}{c}
u\\
v\\
w
\end{array}\right]=\left[\begin{array}{cc}
\boldsymbol{R} & \boldsymbol{t}\\
\boldsymbol{0} & 1
\end{array}\right]\left[\begin{array}{c}
x\\
y\\
z\\
1
\end{array}\right]\\
=\left[\begin{array}{cccc}
R_{11} & R_{12} & R_{13} & u_{0}\\
R_{21} & R_{22} & R_{23} & v_{0}\\
R_{31} & R_{32} & R_{33} & w_{0}\\
0 & 0 & 0 & 1
\end{array}\right]\left[\begin{array}{c}
x\\
y\\
z\\
1
\end{array}\right]
\end{array}\label{eq:extrinsic}
\end{equation}

\end{doublespace}

where$R\in R^{3\times3}$ is the rotation matrix and $t\in R^{3\times1}$
is the translation vector. Since it is a linear projection and both
coordinates share the same unit (meter), the rotation matrix R equals
a multiplication of three sub-rotation matrix $R_{roll}$, $R_{pitch}$
and $R_{yaw}$, as shown in (\ref{eq:rotation}).

\begin{spacing}{2.3}
\begin{equation}
\begin{array}{c}
R=R_{roll}R_{pitch}R_{yaw}\\
R_{roll}=\left[\begin{array}{ccc}
1 & 0 & 0\\
0 & cos(\alpha) & -sin(\alpha)\\
0 & sin(\alpha) & cos(\alpha)
\end{array}\right]\\
R_{pitch}=\left[\begin{array}{ccc}
cos(\beta) & 0 & sin(\beta)\\
0 & 1 & 0\\
-sin(\beta) & 0 & cos(\beta)
\end{array}\right]\\
R_{yaw}=\left[\begin{array}{ccc}
cos(\gamma) & -sin(\gamma) & 0\\
sin(\gamma) & cos(\gamma) & 0\\
0 & 0 & 1
\end{array}\right]
\end{array}\label{eq:rotation}
\end{equation}

\end{spacing}

Where $\alpha$, $\beta$, and $\gamma$ are the rotation angle along
the $x$, $y$ and $z$ axis. The goal of the extrinsic matrix calibration
is to estimate the 6 parameters $(\alpha,\beta,\gamma,u_{0},v_{0},w_{0})$.

The second part is the intrinsic transformation that projects the
3D points in camera coordinate to the 2D image plane. Pinhole camera
model and fisheye camera model are two popular intrinsic models. The
pinhole model is described in (\ref{eq:pinhole}).

\begin{equation}
\left[\begin{array}{c}
i\\
j
\end{array}\right]=\left[\begin{array}{ccc}
f_{x}/w & 0 & i_{0}\\
0 & f_{y}/w & j_{0}
\end{array}\right]\left[\begin{array}{c}
u\\
v\\
1
\end{array}\right]\label{eq:pinhole}
\end{equation}

Where $f_{x}$ and $f_{y}$ are the focus lengths of the lens system
along x and y axis, and $i_{0}$, $j_{0}$ are the offsets on the
target image plane. For the fisheye model, lens distortion and tangential
distortion and skew are also considerated as in (\ref{eq:fisheye}).

\begin{doublespace}
\begin{equation}
\begin{array}{c}
x_{d}=(1+k_{1}r^{2}+k_{2}r^{4}+k_{5}r^{6})\left[\begin{array}{c}
u/w\\
v/w
\end{array}\right]\\
dx=\left[\begin{array}{c}
2k_{3}uv+k_{4}(r^{2}+2u^{2})\\
k_{3}(r^{2}+2v^{2})+2k_{4}uv
\end{array}\right]\\
\left[\begin{array}{c}
i\\
j
\end{array}\right]=\left[\begin{array}{ccc}
f_{x} & \alpha_{c}\cdot f_{x} & i_{0}\\
0 & f_{y} & j_{0}
\end{array}\right]\left[\begin{array}{c}
u\\
v\\
1
\end{array}\right]
\end{array}Recentprogressinautomateddrivingsystem(ADS)andadvanceddriverassistantsystem(ADAS)hasshownthatthecombineduseof3Dlightdetectionandranging(LiDAR)andcameraisessentialforanintelligentvehicletoperceiveitssurroundings.LiDAR-camerafusionrequirespreciseintrinsicandextrinsiccalibrationsbetweenthesensors.However,weobservedthatexistingmethodsperformpoorlyindrivingscenarios.Duetothelimitationofcalibrationequipmentandsusceptibilitytonoise,objectivefunctionsofexistingmethodstendtofailinlong-range3D-2Dprojections.Inthispaper,weintroducedaninteractiveLiDARtocameracalibrationtoolboxtocalibrationtheintrinsicandextrinsictransforms.ThistoolboxautomaticallydetectsthecornerofaplanerboardfromasequenceofLiDARframesandprovidesacontinentuserinterfaceforannotationsoncorrespondingcameraframes.Sincethetoolboxonlydetectsthetopcorneroftheboard,thereisnoneedtoprepareaprecisepolygonplanarboardorcheckerboardwithdifferentreflectivityasexistingmethodsrequire.Furthermore,thetoolboxusesgeneticalgorithmstoestimatethetransformsanditsupportsmultiplecameramodelssuchasthepinholecameramodelandthefisheyecameramodel.ExperimentsusingVelodyneVLP-16LiDARandPointGreyChameleon3camerashowrobustresultsinvariousdrivingscenarios.\label{eq:fisheye}
\end{equation}

\end{doublespace}

Where $r=\sqrt{u^{2}+v^{2}}$.

In general, an intrinsic transformation using pinhole model has 4
parameters $(f_{x},f_{y},i_{0},j_{0})$ and the one using fisheye
has 10 parameters $(f_{x},f_{y},i_{0},j_{0},\alpha_{c,}k_{1},k_{2},k_{3},k_{4},k_{5})$.

\subsection{data collection and processing}

The key issue of point-to-point based approaches in LiDAR to camera
calibration is to collect the point-to-point pairs between LiDAR and
camera coordinates. Since LiDAR point cloud is sparse on the vertical
axis, it is difficult to get the target correspondence on a checkerboard
in a single frame. In our approach, we detect the vertices of the
checkerboard in a sequence of LiDAR frames. 

\begin{figure}
\begin{tabular}{c}
\includegraphics[width=8cm]{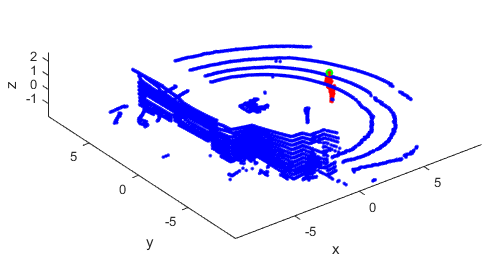}\tabularnewline
(a)\tabularnewline
\includegraphics[width=8cm]{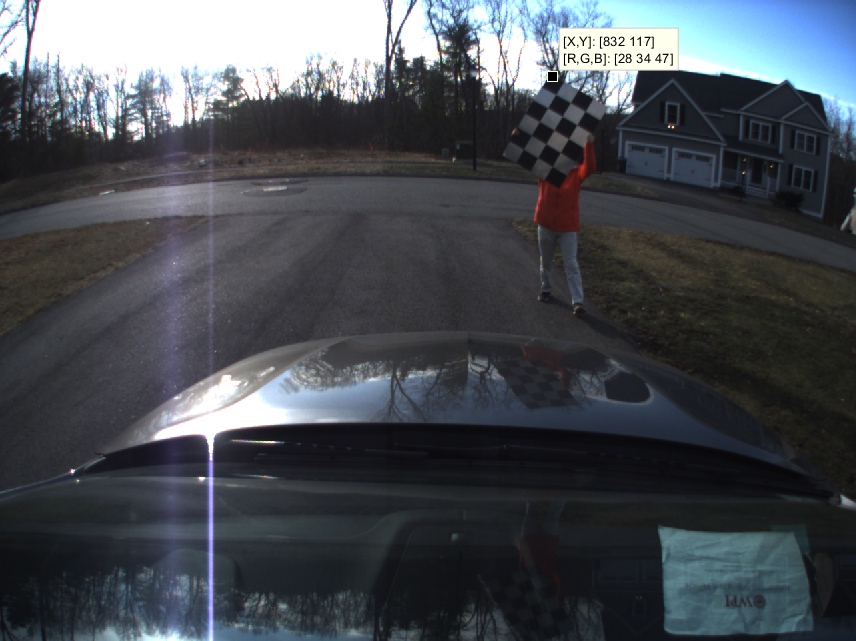}\tabularnewline
(b)\tabularnewline
\end{tabular}

\caption{Example LiDAR frame (a) and example camera frame (b)}
\end{figure}

The proposed approach has the following assumptions: (1) only the
checkerboard and the tester holding the checkerboard exist in the
region of interest (ROI), and (2) the target vertex of the checkerboard
is the highest point in the ROI. Comparing to other related works,
our approach does not require a perfect checkerboard nor a considerable
number of scans of the checkerboards. The proposed approach not only
reduces the difficulty of calibration task but also makes it capable
of calibrating for low-resolution LiDAR devices such as 4-line LiDAR
and 16-line LiDAR. The procedures to detect the vertex in a LiDAR
frame sequence are described in Algorithm \ref{alg:vertex-detection-in}.

\begin{algorithm}
\textbf{Input:} LiDAR Frame Sequence $\{P\}$, RegioLiDAR Frame Sequence
$\{P\}$, Region of interest $ROI$n of interest $ROI$

$\;$1: \textbf{for} LiDAR frame $P_{i}$ in $\{P\}$ \textbf{do}

$\;$2: $\quad$Get point cloud in the $ROI$: $P_{i}\leftarrow\{p|p\epsilon P_{i}\cap ROI\}$

$\;$3: $\quad$\textbf{if} max\_line($P_{i}$ ) - min\_line($P_{i}$
) $\geqslant$3 \textbf{then}

$\;$4: $\qquad$$\hat{P_{i}}$$\leftarrow$ ${p|line(p)=max\_line(P_{i})}$

$\;$5: $\qquad$\textbf{if} Count($\hat{P_{i}}$)=1 \textbf{then}

$\;$6: $\qquad$$\quad$\textbf{return}$\hat{P_{i}}$

$\;$7: $\qquad$\textbf{else then}

$\;$8: $\qquad$$\quad$drop frame $P_{i}$

$\;$9: $\qquad$\textbf{end if}

10: $\quad$\textbf{end if}

11: \textbf{end for}

\caption{vertex detection in LiDAR frame sequence\label{alg:vertex-detection-in}}
\end{algorithm}

When the toolbox starts, it shows a LiDAR frame and the user needs
to input the ROI threshold. For the subsequent LiDAR frames, the toolbox
segments the points inside the ROI, validates the size of segmented
object and searches for the top line inside the ROI. If the top line
has only one point, it is stored as a vertex points, otherwise the
LiDAR frame is dropped. For each frame in which the vertex is detected,
the toolbox asks the user\textquoteright s confirmation and presents
the corresponding camera frame for annotation if a valid vertex is
detected. In practice, the LiDAR and camera sensors might not be perfectly
synchronized, so the toolbox select the camera frame that is captured
nearly at the same time of the LiDAR scan. After each annotation,
the toolbox gets a new point-to-point correspondence between the LiDAR
and camera coordinate.

\subsection{Genetic Algorithm}

The proposed toolbox calibrates the extrinsic and intrinsic transformations
of LiDAR to camera through the genetic algorithm (GA). GA is a widely
used algorithm in parameter estimation since it is model-based, data-driven
and robust to non-linear optimization. In theory, a GA instance rolls
out thousands of alternative parameter sets and selects certain candidates
with least offset as seeds of the next generation. The algorithm terminates
when determined generations are reached or the convergence goal is
reached. Comparing to gradient-based algorithms, GA is capable to
avoid local optima.

In this toolbox, we employe the MATLAB genetic algorithm toolbox as
the GA solver. According to the calibration model, we set 10 parameters
for pinhole-model based solution and 16 parameters for fisheye-model
based solution. The average offset of labeled points projected from
LiDAR to the image plane is selected as the target loss of GA. The
GA process is iterated multiple times until the loss does not decrease.
In each iteration, the toolbox initializes 5 GA slots to avoid local
optima. Each GA slot keeps an 800 population and runs 30 generations
to optimize the parameters. In the first iteration, the upper bound
and lower bound of GA slots are manually determined. During the rest
of the iterations, the toolbox narrows the upper bound and lower bound
according to the optima of last iteration and the number of iterations
it has executed. The iteration algorithm is described in Algorithm
\ref{alg:GA}.

\begin{algorithm}
\textbf{Input:} Initial parameter set $Param$, bounding shift $Bound$,
bounding scale $Scale$

$\;$1: \textbf{for} iteration $Iter$ \textbf{do}

$\;$2: $\quad$Upper Bound $ub\leftarrow Param+Bound$

$\;$3: $\quad$Lower Bound $lb\leftarrow Param-Bound$

$\;$4: $\quad$$\hat{P_{i}}$$\leftarrow$ ${p|line(p)=max\_line(P_{i})}$

$\;$5: $\quad$\textbf{for} slot $Slot$ \textbf{do}

$\;$6: $\qquad$$Param_{i},error_{i}\leftarrow GA(Param,ub,lb)$

$\;$7: $\qquad$$Param\leftarrow\{Param_{i}|error_{i}=min(error)\}$

$\;$8: $\qquad$$Bound\leftarrow Scale\times Bound$

$\;$9: $\quad$\textbf{end for}

10: \textbf{end for}

11: \textbf{return} $Param$

\caption{Genetic Algorithm based transformation parameter optimization\label{alg:GA}}
\end{algorithm}

\section{\textcolor{black}{Experiment\label{sec:Experiment}}}

\begin{figure}
\begin{centering}
\includegraphics[width=8cm]{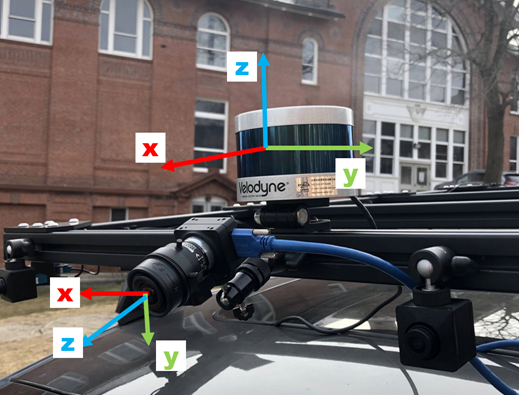}
\par\end{centering}
\caption{LiDAR-camera system on an Lincoln MKZ\label{fig:LiDAR-camera-system setup}}
\end{figure}

\textcolor{black}{In the experiment, we use the proposed toolbox to
calibrate the LiDAR-camera system of our autonomous vehicle prototype.
The system as shown in Figure \ref{fig:LiDAR-camera-system setup}
contains a Velodyne VLP-16 LiDAR and a Point Grey Chameleon 3 camera.
Both devices are installed on the roof rack facing the front. The
LiDAR has 16 row scanners covering a }$[-15\lyxmathsym{\textdegree},15\lyxmathsym{\textdegree}]$\textcolor{black}{{}
vertical field of view and }$360\lyxmathsym{\textdegree}$\textcolor{black}{{}
horizontal one resulting in a }$0.2\lyxmathsym{\textdegree}$\textcolor{black}{{}
horizontal resolution and a }$2\lyxmathsym{\textdegree}$\textcolor{black}{{}
vertical resolution. The camera is adjusted to }$15\lyxmathsym{\textdegree}$\textcolor{black}{{}
looking down to the ground.}

\begin{figure}
\begin{centering}
\includegraphics[width=8cm]{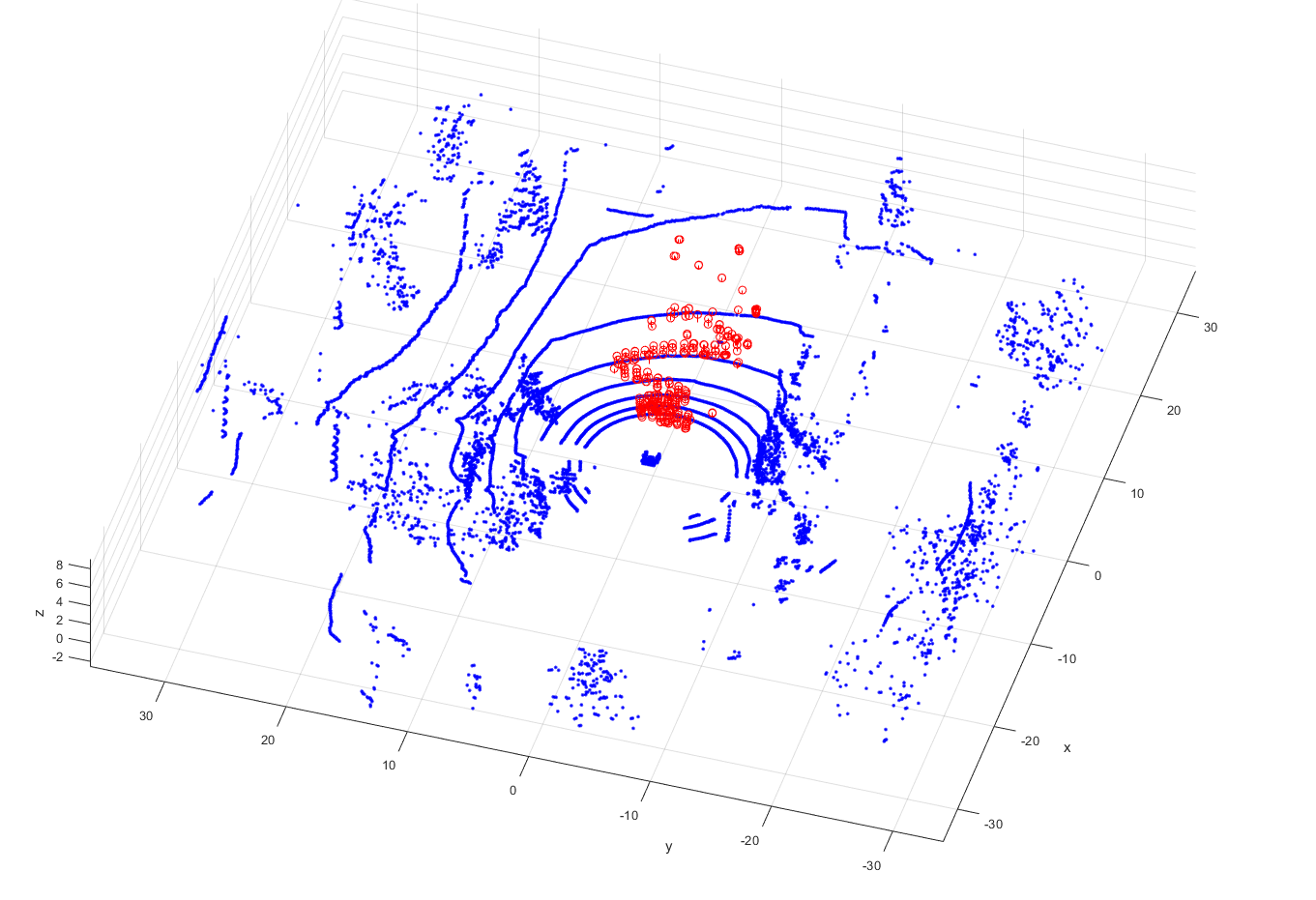}
\par\end{centering}
\caption{Labeled data in LiDAR cooridate frame. The blue dots are the point
cloud from one LiDAR frame, and the stems present the position of
labeled correspondences in the LiDAR coordinate. \label{fig:Labels_vis}}
\end{figure}

\begin{figure}
\begin{centering}
\includegraphics[width=8cm]{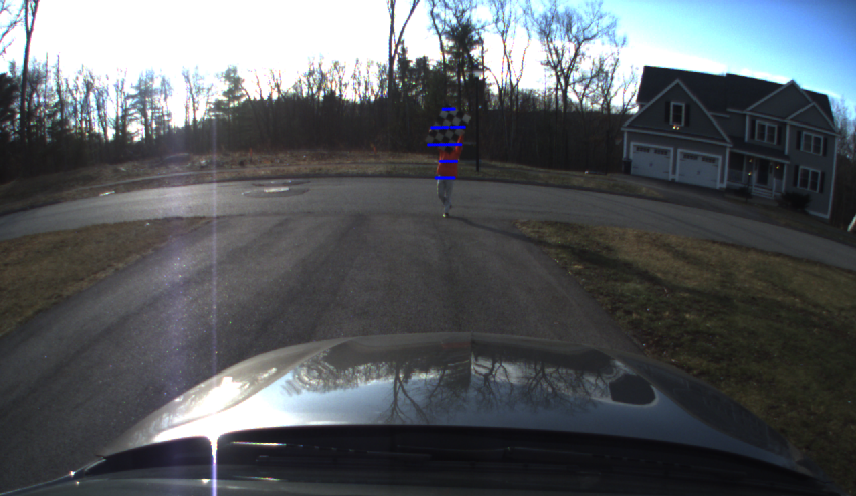}
\par\end{centering}
\caption{Example calibration result \label{fig:Example-calibration-result}}
\end{figure}

\begin{figure}
\begin{centering}
\includegraphics[width=8cm]{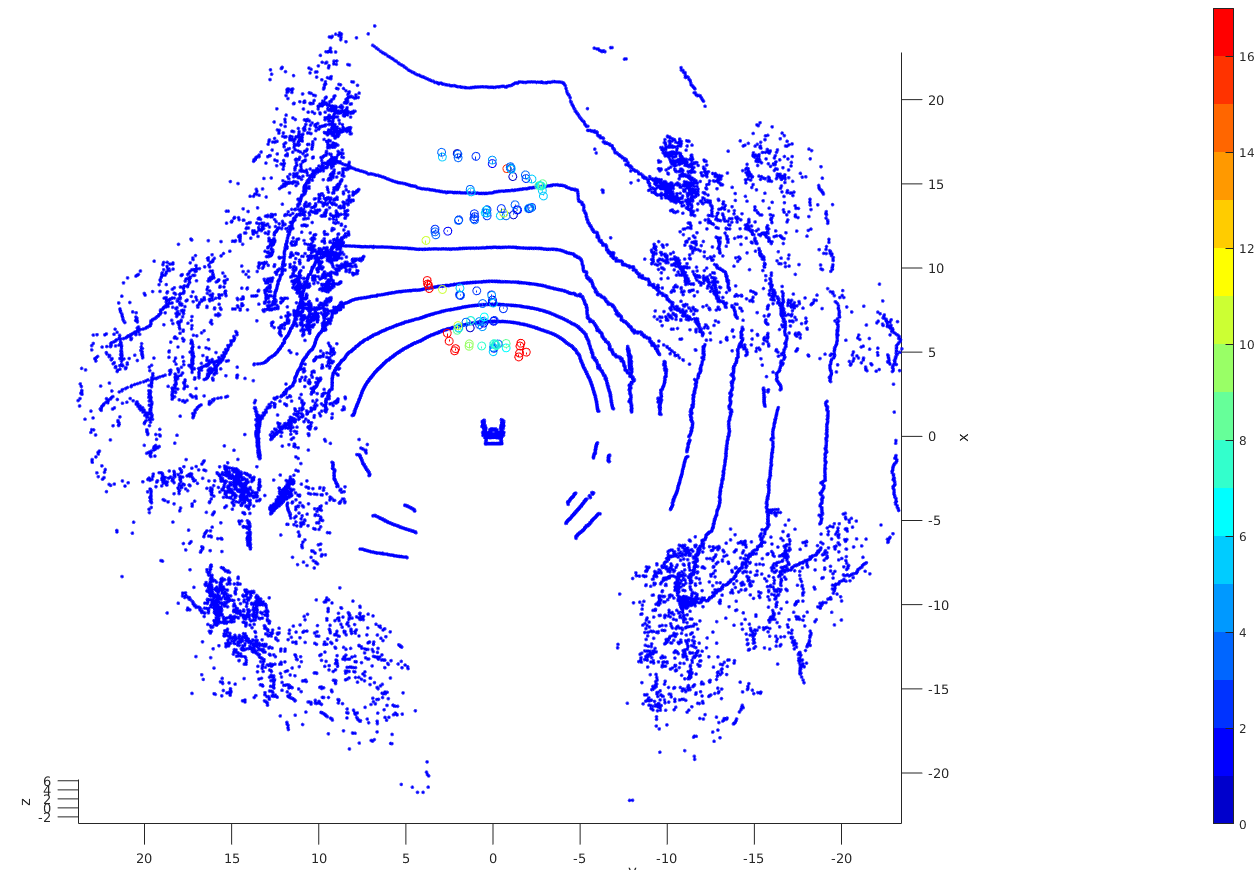}
\par\end{centering}
\caption{Validation of LiDAR-camera calibration using pinhole camera model\label{fig:Validation of calibration pinhole}}
\end{figure}

\begin{figure}
\begin{centering}
\includegraphics[width=8cm]{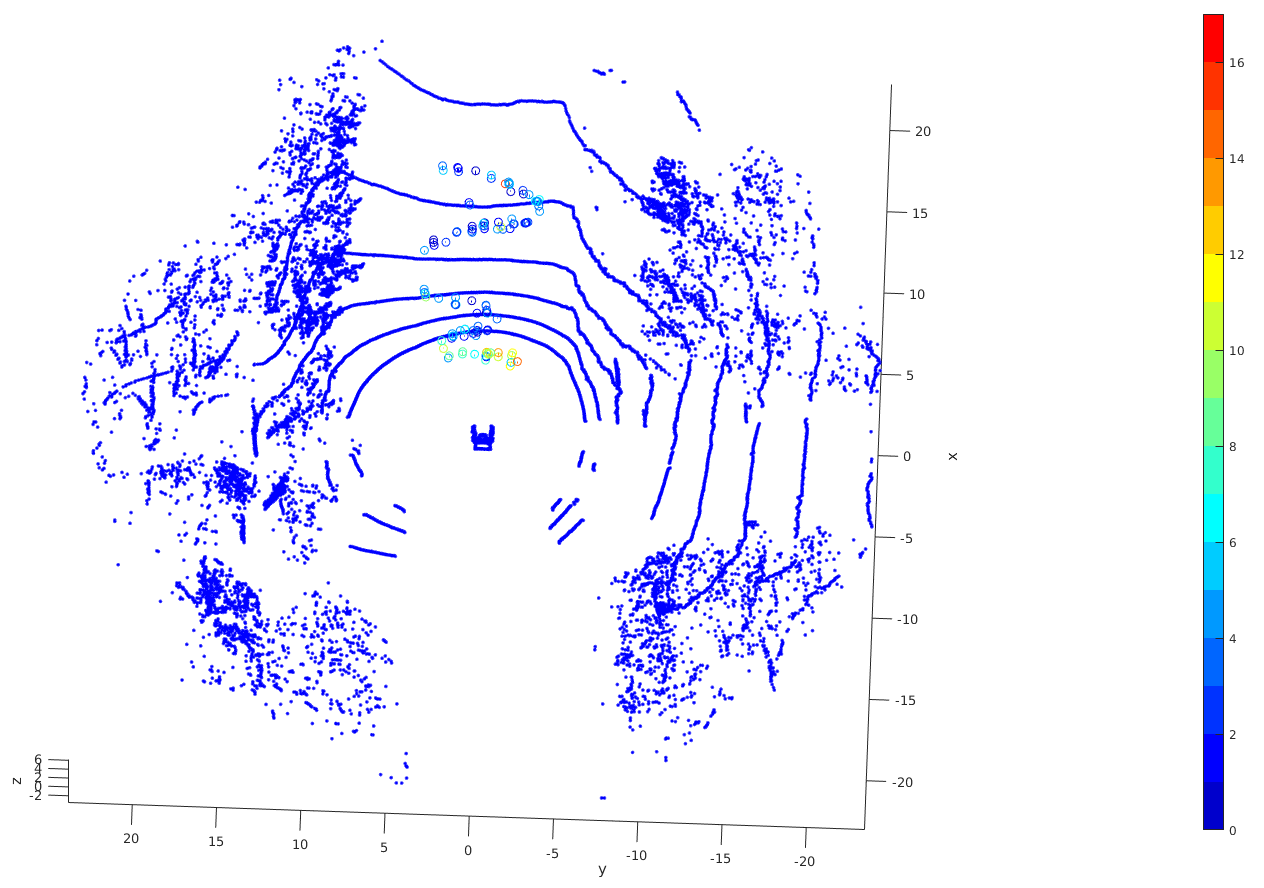}
\par\end{centering}
\caption{Validation of LiDAR-camera calibration using fisheye camera model\label{fig:Validation of calibration fisheye}}
\end{figure}

To calibrate this LiDAR-camera system, we use a $2\times3$ feet checkerboard
rotated $45\lyxmathsym{\textdegree}$ so that one vertex is up. A
person holds the checkerboard and moves around in a $10m\times30m$
field in front of the car. We recorded the data sequence in about
219 seconds that contains 11582 camera frames and 7639 LiDAR frames.
After processing the data, $222$ LiDAR-to-camera correspondences
are collected as shown in Figure \ref{fig:Labels_vis}. The lower
bound and upper bound setting is manually determined as shown in Table
\ref{tab:Boundaries-of-parameters} based on the device setting and
installation. For validation, we record another data sequence with
3293 camera frames and 2177 LiDAR frames, and annotated 358 correspondences.
For the training data, this calibration method is able to achieve
6.35 pixel offset in average for pinhole camera model and 5.02 pixel
for fisheye camera model. On the test data, it results 6.75 pixel
error in average using pinhole camera model and 5.1 pixel using fisheye
camera model. An example projection result is shown in Figure \ref{fig:Example-calibration-result}.
The result shows that the proposed approach generates precise transformation
from LiDAR to camera coordinate. Figure \ref{fig:Validation of calibration pinhole}
and \ref{fig:Validation of calibration fisheye} shows the validation
of the LiDAR-camera transformation along the LiDAR coordinate. It
is obvious that the offset around the camera normal is small, but
as the angle to the camera normal grows, the offset grows to as much
as 16 pixels, which indicates the current camera intrinsic models
may not be precise enough.

\begin{table}
\begin{centering}
\begin{tabular}{|c|c|c|c|c|c|}
\hline 
parameter &
$\alpha$ &
$\beta$ &
$\gamma$ &
$u_{0}$ &
$v_{0}$\tabularnewline
\hline 
lower boundary &
$0.2\pi$ &
$-0.8\pi$ &
$-0.3\pi$ &
$-1$ &
$-1$\tabularnewline
\hline 
upper boundary &
$0.8\pi$ &
$-0.2\pi$ &
$0.3\pi$ &
$+1$ &
$+1$\tabularnewline
\hline 
parameter &
$w_{0}$ &
$f_{x}$ &
$f_{y}$ &
$i_{0}$ &
$j_{0}$\tabularnewline
\hline 
lower boundary &
$-1$ &
$300$ &
$300$ &
$300$ &
$300$\tabularnewline
\hline 
upper boundary &
$+1$ &
$900$ &
$900$ &
$900$ &
$900$\tabularnewline
\hline 
parameter &
$k_{1}$ &
$k_{2}$ &
$k_{3}$ &
$k_{4}$ &
$k_{5}$\tabularnewline
\hline 
lower boundary &
$-1$ &
$-1$ &
$-1$ &
$-1$ &
$-1$\tabularnewline
\hline 
upper boundary &
$+1$ &
$+1$ &
$+1$ &
$+1$ &
$+1$\tabularnewline
\hline 
\end{tabular}
\par\end{centering}
\caption{Boundaries of parameters for GA solver\label{tab:Boundaries-of-parameters}}

\end{table}

\begin{figure}
\begin{centering}
\includegraphics[width=6cm]{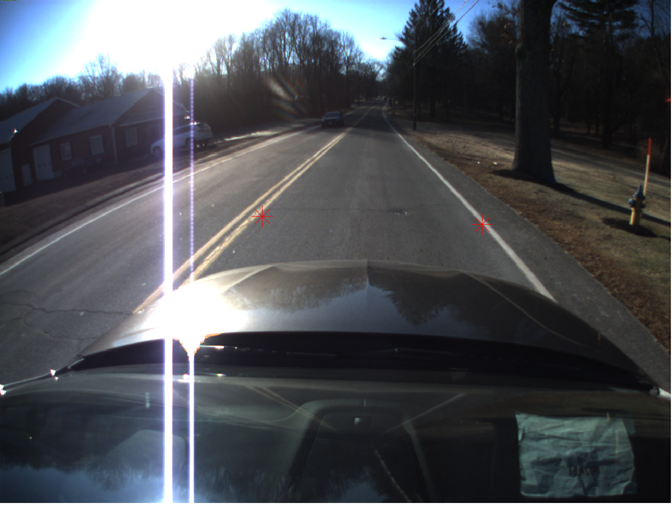}
\par\end{centering}
\begin{centering}
\includegraphics[width=6cm]{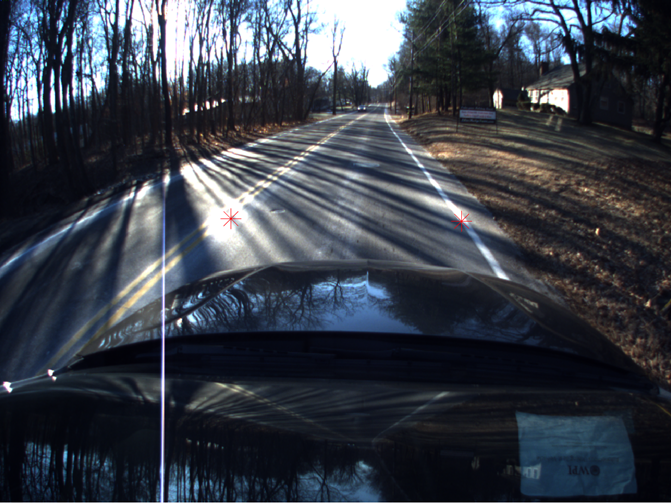}
\par\end{centering}
\begin{centering}
\includegraphics[width=6cm]{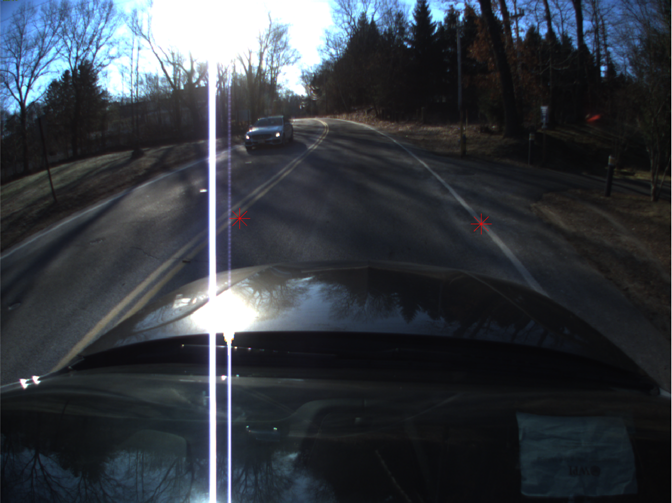}
\par\end{centering}
\caption{Lane markers detected in LiDAR frames and projected to Camera frames\label{fig: Lane marker example}}
\end{figure}

For further testing, we drive the test vehicle on the road and record
a 1043-frame LiDAR and camera sequence. From the data, we detect the
lane markers by thresholding the intensity of LiDAR points in the
lowest ring. We then project the detected lane markers to the corresponding
camera frames using the calibrated transformation function and visualize
them on the camera frames with red asterisks. As shown in Figure \ref{fig: Lane marker example},
the offset of the lane markers is acceptable for automated driving.

\section{\textcolor{black}{Conclusions\label{sec:Conclusions-and-future}}}

\textcolor{black}{In this paper, we propose a novel and accurate approach
for LiDAR-to-camera system calibration. This approach is not sensitive
to the checkerboard quality and works for long-range calibration.
We also develop a MATLAB based toolbox to calibrate the LiDAR and
camera. The toolbox automatically detects the vertex in a LiDAR frame
sequence and provide a convenient user interface to annotate the correspondence
in camera frames. A genetic algorithm based approach is applied to
estimate the extrinsic and intrinsic parameters. Experiment on test
vehicle shows that the toolbox works well on long range calibration
and achieves good results.}

Future work will investigate automatic corner detection on camera
frames, extend the calibration to LiDAR-thermal camera systems, and
investigate the correspondences of 2D to 3D projection.

\textcolor{black}{\bibliographystyle{plain}
\bibliography{12_home_yecheng_IV2019_reference}
}
\end{document}